\documentstyle[epsfig, twocolumn]{esa_sp_latex}

\begin{document}
%

\parindent 0pt
\parskip 10pt plus 1pt minus 1pt
\hoffset=-1.5truecm
\topmargin=-1.0cm
\textwidth 17.1truecm \columnsep 1truecm \columnseprule 0pt 

\hyphenation{ef-fi-cien-cy spec-trom-e-ter spec-tros-co-py e-mis-sion gam-ma
     ar-chiving me-di-um} 

\title{\bf THE GREAT ANNIHILATOR 1E1740.7-2942: MOLECULAR CLOUD CONNECTION
           AND CORONAL STRUCTURE.} 

\author{{\bf  Osmi Vilhu, Diana Hannikainen,
     Panu Muhli and Juhani Huovelin} \vspace{2mm} \\
{Observatory, University of Helsinki, Finland} \vspace{3mm} \\
{\bf Juri Poutanen} \vspace{2mm} \\ 
{Stockholm Observatory, Saltsj\"obaden, Sweden} \vspace{3mm} \\
{\bf Philippe Durouchoux} \vspace{2mm} \\
{DAPNIA, Service d'Astrophysique, CE Saclay, France} \vspace{3mm} \\
{\bf Pierre Wallyn} \vspace{2mm} \\ 
{Jet Propulsion Laboratory, CalTech, USA} } 

\maketitle

\begin{abstract}
Using  $^{12}$CO and $^{13}$CO observations
we present column density maps of the  molecular cloud 
(V$_{LSR}$ = -135 km/s) in the direction of 1E1740.7-2942.
Hydrogen column densities of the cloud  scatter
between N$_H$ = (3.5 - 11)$\times 10^{22}$ cm$^{-2}$,
depending on the method used. From this we conclude,
deriving first a simple analytic formula, that despite of the weakness
of the iron fluorescent 6.4 keV line (Churazov et al. 1996)
the source may
lie  inside the cloud, or at least close to its edge.
The combined ASCA/BATSE
spectrum  from September 1993 and 1994
can be modelled with  a two-phase accretion disc corona model, 
where the  hot region is detached from the cold disc. Geometrically
the  hot phase   can be interpreted e.g.
as a number of active regions (magnetic loops) 
 above the disc, or   as a  spherical hot cloud 
around the central object.    \vspace {5pt} \\

  Keywords: Galactic Center, Black hole candidates, Molecular clouds,
            Gamma-ray spectra, accretion discs

\end{abstract}

\section{INTRODUCTION}

The discovery of a massive molecular cloud (at V$_{LSR}$ $\approx$ -130 km/s)
in the direction of the bright X-ray
source 1E1740.7-2942 (Bally and Leventhal 1991; Mirabel et al. 1991) raised
a picture of a compact source embedded in a molecular cloud.
It was suggested that the unique properties of this object are related to the
presence of dense gas surrounding it. In particular, the Bondi-Hoyle
accretion onto a single black hole was discussed.

The ASCA data imply 
 N$_H$ = 8$\times 10^{22}$ cm$^{-2}$  for the low energy absorption
(Sheth et al. 1996), and twice of that
for the iron K-edge absorption (Churazov et al. 1996).
Based on these column density estimates Churazov et al. concluded
that the weakness of the iron 6.4 keV line suggests that the source is
located behind the cloud.

\section{THE MOLECULAR LINE OBSERVATIONS} 

\label{sec:molec} 

The observations were performed between August 1 - 8, 1993, using the
Swedish ESO Submillimeter Telescope (SEST)
 situated at La Silla in Chile. The front-end was
a Schottky receiver and the 1086 MHz bandwidth acousto-optical spectrometer
(AOS) was used as the back-end. The intensity calibration was performed using
the chopper-wheel method, the background was eliminated with the position
switching mode.
The half power beamwidth (FWHM) of the telescope at 115 GHz is 45 arc sec.

We observed two lines $^{12}$CO(J=1-0) (115 GHz) and $^{13}$CO(J=1-0) (110 GHz)
at 35 spatial positions around 1E1740.7-2942. In both transitions the line
at V$_{LSR}$ = -135 km/s can be clearly identified (Fig.~1)
and the antenna brightness temperature was integrated
between $-115$ km/s and $-150$ km/s to give the total flux F$_{\rm cloud}$.
Three different methods were used to calculate
N$_H$ = 2$\times$N(H$_2$):

\vspace{0.2cm}

{\bf Method 1.} The standard conversion (see Bally and Leventhal 1991; Fig.~2)

\vspace{0.1cm} 
N(H$_2$)/F$_{\rm cloud}$($^{12}$CO) =
     2.6$\times$10$^{20}$ cm$^{-2}$/K km~s$^{-1}$. 

\vspace{0.2cm}

{\bf Method 2.} Scaling with the LTE-relation 
   given by Bally and Leventhal (1991) with T$_{exc}$ = 15 K (see Fig. ~3)

\vspace{0.1cm}

N(H$_2$)/F$_{\rm cloud}$($^{13}$CO) =
    5.1$\times$10$^{20}$ cm$^{-2}$/K km~s$^{-1}$.

\vspace{0.2cm}

{\bf Method 3}. Assuming LTE like in the method 2, but computing T$_{exc}$ and
                $\tau$ with the help of $^{12}$CO
               and $^{13}$CO, and finally using the conversion N(H$_2$) =
   1.0$\times$10$^6$$\times$N($^{13}$CO) (Gahm et al. 1993; see Fig. ~4). 

\vspace{0.2cm}

The column density maps based on these methods are shown in  Figures~2-4. 
In the direction of 1E1740.7-2942, the methods 1, 2 and 3 give values of
N$_H$ = 11, 3.5 and 7, respectively (in units of 10$^{22}$ cm$^{-2}$ and 
in the position of the circle centrum in Figs. 2 - 4).


\input psbox.tex
\begin{figure}[h]
  \begin{center}
    \leavevmode
\mbox{\psboxto(8cm;0cm){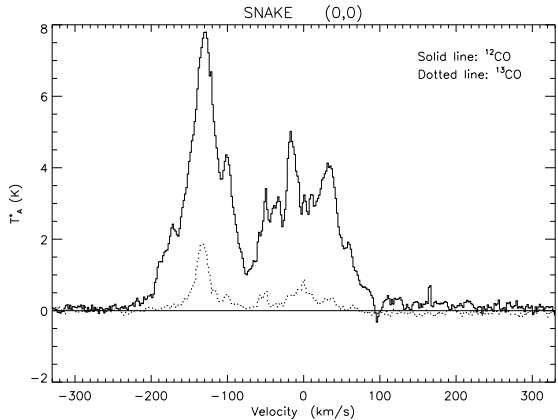}}
  \end{center}
  \caption{\em $^{12}$CO (solid) and $^{13}$CO (dotted) spectra at the
         (0,0) position of the maps in Figs.~2-4. } 
  \label{fig:1} 

\input psbox.tex
\end{figure}
\begin{figure}[h]
  \begin{center}
    \leavevmode
\mbox{\psboxto(10cm;0cm){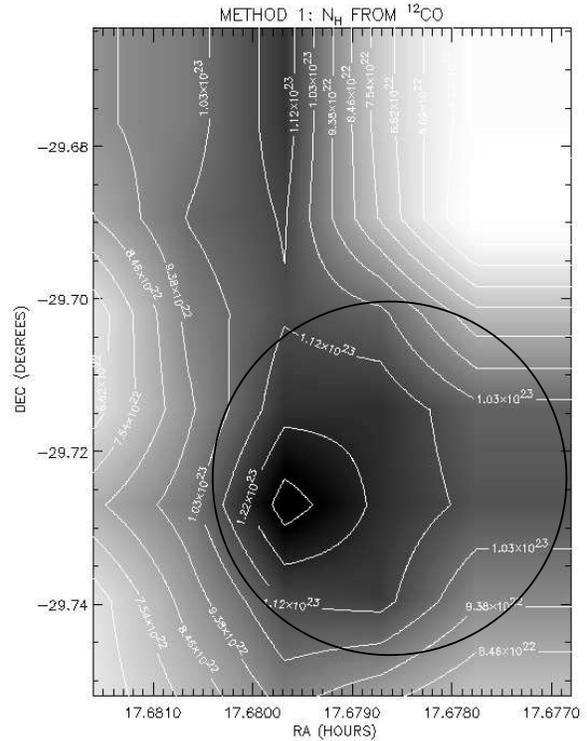}}
 \end{center}
  \caption{\em The Hydrogen column density map around 1E1740.7-2942 from
the $^{12}$CO line observations with a linear scaling (method 1,
see the text). The
ASCA resolution  circle (PSF diameter 2.9 arc min = 7 pc at 8.5 kpc) 
centered at the  
      radio-jets (Mirabel et al. 1992) is shown. The darkest region
corresponds to the largest column density. The range of N$_H$ is
between (9 - 13)$\times$10$^{22}$ cm$^{-2}$ inside the circle. } 
  \label{fig:2} 
\end{figure}

\input psbox.tex
\begin{figure}[h]
  \begin{center}
    \leavevmode
\mbox{\psboxto(10cm;0cm){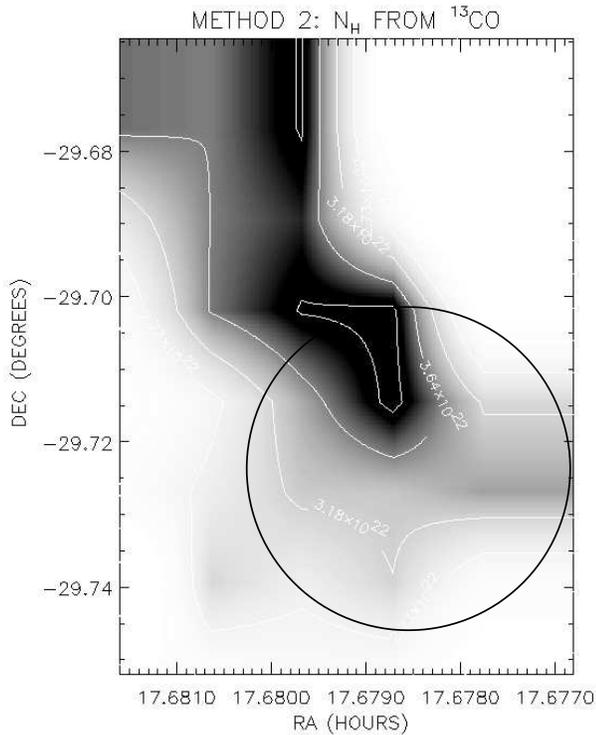}}
  \end{center}
  \caption{\em As in Fig.~2, but from the $^{13}$CO line observations with
a linear scaling (method 2, see the text). The darkest regions
correspond to the largest column density. The range of N$_H$ is between
(3 - 4)$\times$10$^{22}$ cm$^{-2}$ inside the circle. } 
  \label{fig:3} 
\end{figure}

\input psbox.tex
\begin{figure}[h]
  \begin{center}
    \leavevmode
\mbox{\psboxto(10cm;0cm){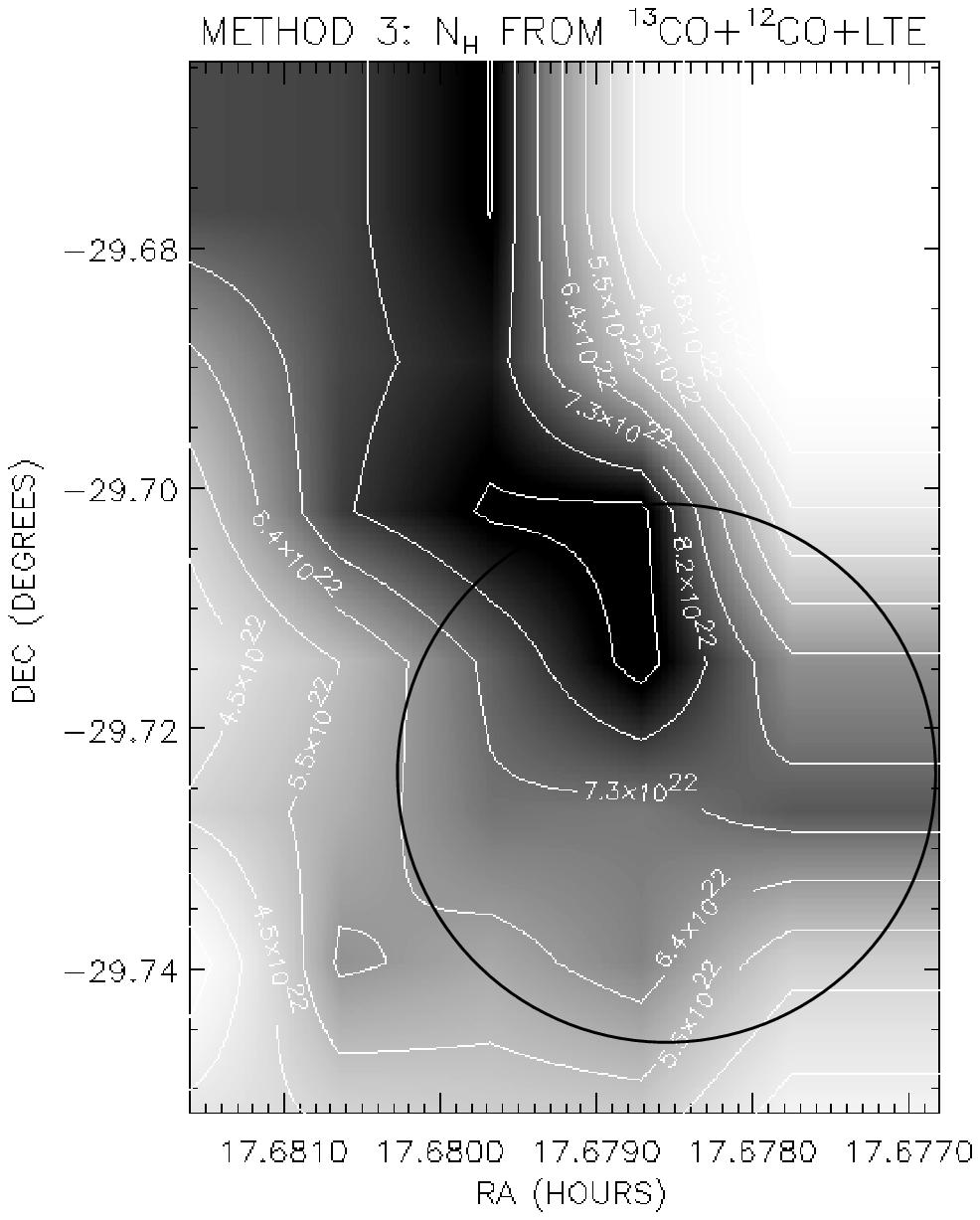}}
  \end{center}
  \caption{\em As in Figs.~2 and 3, but from the LTE method using $^{12}$CO
 and $^{13}$CO (method 3, see the text). The darkest regions
correspond to the largest column density. The range of N$_H$ is between
(5 - 9)$\times$10$^{22}$ cm$^{-2}$ inside the circle.} 
  \label{fig:4} 
\end{figure}

\section{IS 1E1740.7-2942 INSIDE THE CLOUD?}

Churazov, Gilfanov and Sunyaev (1996) showed that
in the ASCA data there is no (or very weak)
evidence of the fluorescent iron line at 6.4 keV. The line equivalent width
is smaller than 20 eV with 90 per cent confidence, the 'best value' being
around 10 eV.
Further, Churazov et al. showed that the  column must be less than
(2 - 3)$\times$10$^{22}$ cm$^{-2}$, if the source is in the center 
of the cloud. They concluded that 1E1740.7-2942 is behind the  cloud. 

Our  measurements  in the direction of 1E1740.7-2942
N$_H$ = (3.5 -11)$\times$10$^{22}$ cm$^{-2}$ suggest that the source location
may be  behind the cloud to explain both the iron K-edge absorption and the
standard IS-absorption in this general direction
(5-7)$\times$10$^{22}$ cm$^{-2}$ (Sheth et al. 1996).
However, since the total column (integrating over all velocities in Fig.~1)
is large (6$\times$10$^{23}$ cm$^{-2}$), 1E1740.7-2942 may equally well lie
in front of the cloud.

Assume that 1E1740.7-2942 is at a distance
$d$ from the centre of a spherical
homogeneous cloud with radius $r$.
For a Thomson optically thin cloud,  
it is rather simple algebra to derive
the following relation between the equivalent width EW of the fluorescent line,
$a\equiv d/r$, the column density across the cloud center N$_H$   
and iron abundance NFe (number density):
\begin{equation}
\label{eq:EW} 
EW= 30\; N_{H,23}\;  (NFe/NFe_\odot) \;  G(a)\; , 
\end{equation} 
where the equivalent width EW is in units of eV, 
column density is in units of $10^{23}$cm$^{-2}$, and the geometrical
factor $G(a)$ is given by the formula: 
\begin{equation}
G(a) = \frac{1}{2}\left[ 1+ \frac{1-a^2}{2a}\ln\frac{1+a}{\vert 1-a\vert} \right].
\end{equation} 
To derive the numerical factor in equation~1 we used the 
same values 
for the absorption cross section at the iron K-edge 
(3.5$\times 10^{-20}$ cm$^2$ per iron atom )
and for the fluorescent yield  ($W = 0.35$) as 
Churazov et al. (1996) did.

For a solar abundance of iron (NFe/NH = 10$^{-4.5}$) , this equation  
predicts $a=1.1$ (i.e. the source is  close to the edge of the
cloud) for the largest column
N$_{H}$ = 11$\times$10$^{22}$ cm$^{-2}$  (method 1) and
for the 'best' EW = 10 eV derived by Churazov et al..
The smallest column from the method 2 (3.5$\times$10$^{22}$)  gives
$a = 0.1$ (i.e. the source is close to the center of the cloud). 
Values of $a$  range between 0.9 and 1.5 for $NFe=2NFe_\odot$ which can not
be rejected due to the large observed iron K-edge. 

The real column of the cloud
might be somewhat different from the values used. 
However,  the size of the scattering
induced area inside the ASCA resolution circle is small enough
to make our estimates reasonably good (see Figs. 2 - 4, extreme values
of N$_H$ range between (3 - 13)$\times$10$^{22}$ cm$^{-2}$ inside the circles).
 After all, the largest uncertainties in
column density estimates come from the method used.

\section{ASCA/BATSE SPECTRAL MODELLING} 

 We attempted
 spectral modelling using  two-phase disc-corona models described
by Poutanen and Svensson (1996). 
In these models, soft radiation from the accretion disc gets comptonized
by hot electron (-positron) gas in the corona. 
 Different coronal geometries can be considered: slab, hemisphere, and
cylinder. 
The effects of different viewing angles
 and separation of the  hot  active regions from the disc 
(covering factor $g$) can also be modelled.
This is particularly relevant to 1E1740.7-2942,
since the weakness
of the iron line suggests that the reflection component is small. This means
that the disc is viewed edge-on 
and/or the cold disc covers much less than $2\pi$ solid angle as viewed from
the X-ray source. 

The models with large  $g$ 
tend to have too small
fluxes at BATSE energies. To compensate this one can decrease the
number of reprocessed soft photons entering the corona (small $g$-factor).  
By the energy balance this
increases the optical depth, allowing more scattering orders
at BATSE energies. Geometrically small $g$ means that the
hot corona is {\bf detached from the disc} ({\bf like} apexes of {\bf magnetic 
coronal loops}) {\bf or} is {\bf situated 
in a central  hole} inside the classical disc (Shapiro et al. 1976,
Narayan 1996).

We used as  input observations those
from  ASCA  (Sheth et al. 1996) and from
 the simultaneous BATSE archive observations analyzed by the
 JPL earth occultation
 method (Wallyn et al. 1996). We do not attempt to give 'the best fit'
 model, just demonstrate one possible model.

 Figure~5 shows the average ASCA spectrum (September 26, 1993 and
 September 9-12, 1994; Sheth et al. 1996) fitted well with a power-law
 fit ($\alpha$ = 1.1 and N$_H$ = 8 $\times 10^{22}$ cm$^{-2}$),
 together with the BATSE data averaged over 13 days around the ASCA
 observing dates.
 In addition, the  BATSE standard state spectrum is shown 
(high state minus low state in
the 1989 - 91 data to subtract the apparent background contamination). The
spectrum obtained in this way is almost identical to the 
SIGMA standard
 state spectrum (Sunyaev et al. 1991). 
The high energy  rise  can not be
modelled with thermal models. It needs a separate physical component,
and part
 of that may come from a possible diffuse radiation in the BATSE field
of view.

\input psbox.tex
\begin{figure}[h]
  \begin{center}
    \leavevmode
\mbox{\psboxto(8cm;0cm){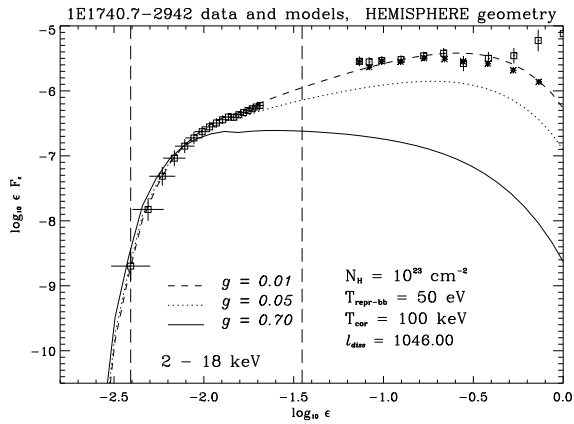}}
 \end{center}
  \caption{\em The average ASCA spectrum of 1E1740.7-2942
from September 26, 1993, and September 9-12, 1994 (Sheth et al. 1996),
together with the BATSE spectrum averaged over 13 days around these
dates (data points with error bars). 
The standard state BATSE spectrum (almost identical to
 the standard SIGMA spectrum by Sunyaev et al. 1991) is also shown (stars).
Edge-on
models with  no reflection component and with detached active regions
are overplotted (g = 0.05 and 0.01). For comparison, the
same model with active regions in touch with the disc is shown by the
solid line ($g = 0.7$). The x-axis is log$_{10}\epsilon$ where
$\epsilon$ = E(keV)/511 and the y-axis equals to 
log$_{10}(\epsilon^2$F$_{ph}$) where F$_{ph}$ is the observed 
photon flux (ph/cm$^{2}$/sec/keV).} 
  \label{fig:5} 
\end{figure}

 Models with  hemisphere type active regions are
 overplotted in Figure~5. They  have  coronal temperature
 T$_{cor}$ = 100 keV and soft photons black body temperature T$_{bb}$ = 50 eV.
The models are edge-on and 
the reflection component was switched off due to the weakness of the iron 
line. 
Different geometrical factors were used 
(e.g. $g = 0.01$ corresponds to the case when $1\%$ of the radiation 
 reprocessed and reflected from the cold disc enters back the active region).
The optical depths for the models with g = 0.7, 0.05 and 0.01 are 1.05, 2.8 
and 3.75, respectively. 

In the case of electron-positron pair dominated 
corona,
 the compactness  parameter
$l_{diss} = (L/H)(\sigma_T/m_ec^3)$ can be used
 to estimate the dimension H of the active region.
 Assuming the luminosity $L = 3\times 10^{37}$ ergs/s for the
 standard state (Sunyaev et al. 1991), one obtains for the best model in Fig. 5
with  g = 0.01 (l$_{diss}$ = 1046)
$H = 8\times 10^{5}$ cm.
This corresponds to 3R$_{Sch}$ for a solar mass central object. 
This type of central hot flow solutions were found already by Shapiro et al. 
(1976). 
For hotter T$_{cor}$ this dimension increases
to match better with a more massive BH.
Assuming many
 active regions, H scales inversely with the number.
 For $e^+e^-$-pairs in the active region, the number density
can be estimated in the model as
$n = 1.5\times 10^{24} \tau /H \approx 7\times 10^{18} cm^{-3}$.



\end{document}